\RequirePackage{fix-cm} 
\documentclass[a4paper, twoside, 10pt, dvips, reqno]{amsart}

\usepackage{fixltx2e}     

\usepackage{amssymb}
\usepackage{amsmath}

\usepackage{a4wide}

\usepackage{indentfirst}
\usepackage{graphicx}
\usepackage{psfrag}

\usepackage[usenames,dvipsnames]{pstricks}
\usepackage{epsfig}
\usepackage{pst-grad} 
\usepackage{pst-plot} 

\usepackage{subfigure}

\usepackage{longtable}
\usepackage{ctable, caption}

\usepackage[english]{babel}
\usepackage[latin1]{inputenc}

\usepackage{color}
\definecolor{oneblue}{rgb}{0,0.0,0.75}
\usepackage[colorlinks,
            urlcolor=oneblue,
            linkcolor=oneblue,
            citecolor=oneblue,
            bookmarksopen=false,
            pagebackref]{hyperref}

\numberwithin{equation}{section}

\newtheorem{rem}{Remark}

\newcommand{\R}{\mathbb{R}}

\newcommand{\Ah}{\mathcal{A}}

\newcommand{\Fh}{\mathcal{F}}
\newcommand{\Gh}{\mathcal{G}}
\newcommand{\Hh}{\mathcal{H}}

\newcommand{\Fip}{\Fh_{i+\frac12}}
\newcommand{\Fim}{\Fh_{i-\frac12}}
\newcommand{\Gip}{\Gh_{i+\frac12}}
\newcommand{\Gim}{\Gh_{i-\frac12}}
\newcommand{\dx}{\Delta x}
\newcommand{\dt}{\Delta t}

\makeatletter
\newcommand{\sech}{\mathop{\operator@font sech}}
\newcommand{\sign}{\mathop{\operator@font sign}}
\makeatother

\newcommand{\diag}{\mathop{\mathrm{diag}}}

\usepackage{acronym}
\acrodef{FV}{Finite Volumes}
\acrodef{FD}{Finite Differences}
\acrodef{DG}{Discontinuous Galerkin}
\acrodef{PDEs}{Partial Differential Equations}
\acrodef{SWs}{Solitary Waves}
\acrodef{NSWE}{Nonlinear Shallow Water Equations}

\begin{document}

\title{Finite volume schemes for Boussinesq type equations}

\author[D. Dutykh]{Denys Dutykh$^*$}
\address{LAMA UMR 5127, Universit\'e de Savoie, CNRS, Campus Scientifique, 73376 Le Bourget-du-Lac France}
\email{Denys.Dutykh@univ-savoie.fr}
\urladdr{http://www.lama.univ-savoie.fr/~dutykh/}
\thanks{$^*$ Corresponding author}

\author[T. Katsaounis]{Theodoros Katsaounis}
\address{Department of Applied Mathematics, University of Crete, Heraklion, 71409 Greece}
\email{thodoros@tem.uoc.gr}
\urladdr{http://www.tem.uoc.gr/~thodoros/}

\author[D. Mitsotakis]{Dimitrios Mitsotakis}
\address{UMR de Math\'ematiques, Universit\'e de Paris-Sud, B\^atiment 425, P.O. Box, 91405 Orsay, France}
\email{Dimitrios.Mitsotakis@math.u-psud.fr}
\urladdr{http://sites.google.com/site/dmitsot/}

\begin{abstract}
Finite volume schemes are commonly used to construct approximate solutions to conservation laws. In this study we extend the framework of the finite volume methods to dispersive water wave models, in particular to Boussinesq type systems. We focus mainly on the application of the method to bidirectional nonlinear, dispersive wave propagation in one space dimension. Special emphasis is given to important nonlinear phenomena such as solitary waves interactions.
\end{abstract}

\keywords{finite volume method; dispersive waves; solitary waves; runup; water waves}

\maketitle

\section{Introduction}\label{sec:intro}

The simulation of water waves in realistic and complex environments is a very challenging problem. Most of the applications arise from the areas of coastal and naval engineering, but also from natural hazards assessment. In this work we will study numerically  bidirectional water wave models. Specifically, we consider the following family of Boussinesq type systems of water wave theory, introduced in \cite{BCS},  written in nondimensional, unscaled variables
\begin{equation}\label{E1.4i}
\begin{aligned}
& \eta_t+u_x+(\eta u)_x+a\, u_{xxx}-b\,\eta_{xxt}=0,\\
& u_t+\eta_x+uu_x+c\,\eta_{xxx}-d\,u_{xxt}=0,
\end{aligned}
\end{equation}
where $a,\, b, \, c, \,d\in\R$, $\eta=\eta(x,t)$, $u=u(x,t)$ are real functions defined for $x\in \mathbb{R}$ and $t\geq 0$.
\begin{equation*}\label{E1.5}
a=\frac{1}{2}(\theta^2-\frac{1}{3})\nu,\ 
b=\frac{1}{2}(\theta^2-\frac{1}{3})(1-\nu),\ 
c=\frac{1}{2}(1-\theta^2)\mu,\ 
d=\frac{1}{2}(1-\theta^2)(1-\mu),
\end{equation*}
where $0\le\theta\le 1$ and $\mu, \nu\in\R$.

Finite volume method is well known for its accuracy, efficiency and robustness for approximating solutions to conservation laws and in particular to nonlinear shallow water equations. The aforementioned bidirectional models \eqref{E1.4i} are rewritten in a conservative form and discretization by the finite volume method follows. Three different numerical fluxes are employed
\begin{itemize}
\item a simple \emph{average flux} (m-scheme),
\item a \emph{central flux}, (KT-scheme) \cite{NT, KT}, as a representative of central schemes,
\item  a \emph{characteristic flux} (CF-scheme), as a representative of the linearized Riemann solvers, \cite{Ghidaglia1996}.
\end{itemize}
along with \emph{TVD}, \emph{UNO} and \emph{WENO} reconstruction techniques, \cite{Sweby1984, HaOs, LOC}. Time discretization is based on Runge-Kutta (RK) methods which preserve the total variation diminishing (TVD) property of the finite volume scheme, \cite{Spiteri2002}. We use explicit RK methods since we work with BBM type systems \eqref{E1.4i} and not with KdV equation which is well known to be notoriously stiff.

The present text is organized as follows. In Section \ref{sec:intro} we present the mathematical model under consideration and the context of this study. Then, Section \ref{sec:num} contains a brief description of various numerical schemes we implemented. Accuracy tests and several numerical results on head-on collision of solitary waves are presented in Section \ref{sec:numres}. Finally, some conclusions of this study are outlined in Section \ref{sec:concl}.

\section{Numerical schemes}\label{sec:num}

In the present section we generalize the finite volume method to systems \eqref{E1.4i} of dispersive PDEs. Boussinesq system \eqref{E1.4i}  can be rewritten in a conservative like form as follows:
\begin{equation}\label{E2.2}
({\bf I}-{\bf D}){\bf v}_t+\left[{\bf F}({\bf v})\right]_x+\left[{\bf G}({\bf v})\right]_x=0, 
\end{equation}
where ${\bf v}=(\eta,u)^T$, ${\bf F}({\bf v})=((1+\eta)u,\eta+\frac{1}{2}u^2)^T$, ${\bf G}({\bf v})=(a\,u_{xx},c\,\eta_{xx})$, and ${\bf D}=\diag\,(b\,\partial^2_x,d\,\partial^2_x)$.  The simplest discretization is based on the average fluxes $\Fh^m$ for ${\bf F}$ and $\Gh^m$ for ${\bf G}$. For the other two choices of the numerical flux $\Fh$ the evaluation of Jacobian is needed.  Let  $A$ denotes the Jacobian of ${\bf F}$, then 
$$
A=\left(
\begin{array}{cc}
u & 1+\eta\\ 
1 & u 
\end{array} \right),
$$ 
with eigenvalues  $\lambda_i=u\pm \sqrt{1+\eta}$, $i=1,2$. It is readily seen, since ${\bf F}$ is a hyperbolic flux, that $A$ can be decomposed as $A=L\Lambda R$ thus for the characteristic flux $\Fh^{CF}$ we have with $\mu=\frac{W+V}{2}$, $s_i=\sign(\lambda_i), \ i=1,2$
$$
\Ah(W,V)=\left(
 \begin{array}{cc} 
 \frac{1}{2}(s_1+s_2) & \frac{1}{2}\sqrt{1+\mu_1}(s_1-s_2)\\ 
\frac{s_1-s_2}{2\sqrt{1+\mu_1}} & \frac{1}{2}(s_1+s_2) 
\end{array} \right).
$$ 
For evaluating the numerical fluxes $\Fh, \ \Gh$ simple cell averages or higher order approximations such as UNO2 or WENO can be used. For more details we refer to our original research article \cite{Dutykh2010}.

\begin{rem}
The discretization of the elliptic operator ${\bf D}$ is based on the standard centered difference. This is a second order accurate approximation and it is compatible with the TVD2 and UNO2 reconstructions. For higher order interpolation we need to modify the elliptic and flux discretization to match the reconstruction's order of approximation.  Indeed, the finite volume scheme is modified as 
\begin{equation*}\label{FV1a}
\frac{d}{dt}\left[\frac{{\bf V}_{i-1}+10{\bf V}_i+{\bf V}_{i+1}}{12}-(b, d)\frac{{\bf V}_{i+1}-2{\bf V}_i+{\bf V}_{i-1}}{\dx^2}\right] + \frac{\Hh_{i-1}+10\Hh_i+\Hh_{i+1}}{12}=0
\end{equation*}
where $\Hh_i=\frac{1}{\dx}(\Fip-\Fim) +\frac{1}{\dx}(\Gip-\Gim)$, is a fourth order accurate approximation. 
\end{rem}

\begin{rem}
In the sequel for the discretization of the dispersive term ${\bf G}$ we use mainly the average numerical flux $\Gh^m$ defined as $\Gh^m_{i+\frac{1}{2}}=(a,c)\frac{{\bf Y}_i+{\bf Y}_{i+1}}{2}$, where ${\bf Y}_i=\frac{{\bf V}_{i+1}-2{\bf V}_i+{\bf V}_{i-1}}{\Delta x^2}$. In case of higher order WENO reconstructions we use the average numerical flux based on the reconstructed values of ${\bf Y}_i$ i.e. the flux $\Gh^{lm}_{i+\frac{1}{2}}=(a,c)\frac{{\bf Y}^L_{i+\frac{1}{2}}+{\bf Y}^R_{i+\frac{1}{2}}}{2}$, where ${\bf Y}^L_{i+\frac{1}{2}} $ and ${\bf Y}^R_{i+\frac{1}{2}}$ are reconstructed values of ${\bf Y}_i$.
\end{rem}

\subsubsection{Boundary conditions}

In the case of Bona-Smith type systems with flat bottom we consider herein only the initial-periodic boundary value problem which is known to be well-posed \cite{ADM1}.

\section{Numerical results}\label{sec:numres}

For the Boussinesq system \eqref{E1.4i} we present first results demonstrating  the accuracy of the finite volume scheme. Then, we study interaction of solitary waves.

\subsection{Accuracy test, validation}

We consider the initial value problem with periodic boundary conditions for the Bona-Smith systems with known solitary wave solutions \cite{Chen1998} to study the accuracy of the finite volume method: 
\begin{equation*}
\begin{array}{l}
\eta(\xi)=\eta_0\, {\sech}^2(\lambda\xi),\\
u(\xi)=B\,\eta(\xi),
\end{array}
\end{equation*}
with
\begin{equation*}
\textstyle{
\begin{array}{cc}
  \eta_0=\frac{9}{2}\cdot\frac{\theta^2-7/9}{1-\theta^2},&
  c_s=\frac{4(\theta^2-2/3)}{\sqrt{2(1-\theta^2)(\theta^2-1/3)}},\\
  \lambda=\frac{1}{2}\sqrt{\frac{3(\theta^2-7/9)}{(\theta^2-1/3)(\theta^2
  -2/3)}},& B=\sqrt{\frac{2(1-\theta^2)}{\theta^2-1/3}}.
  \end{array}}
\end{equation*}
We fix $\theta^2=8/10$ in the system and an analytic solitary wave of amplitude $\eta_0=1/2$ is used as the exact solution in $[-50, 50]$  computed up to $T=100$. The error is measured with respect to discrete $L^2$ and $L^{\infty}$ norms, namely we use:
\begin{align*}
& E_h^2(k)=\|U^k\|_h/\|U^0\|_h, \quad  \|U^k\|_h=\left(\sum_i \dx |U^k_i|^2\right)^{1/2} , \\
& E_h^{\infty}(k)=\|U^k\|_{h,\infty}/\|U^0\|_{h,\infty}, \quad \|U^k\|_{h,\infty}=\max_i |U^k_i|,
\end{align*} 
where $U^k=\{U^k_i\}_i$ denotes the solution of the fully-discrete scheme at the time $t^k=k\, \dt$. The expected theoretical order of convergence was confirmed for all finite volume methods we presented above. Two indicative cases are reported in Table \ref{ROC} for the average flux and TVD2 implementation with MinMod limiter.

\begin{table}%
\centering
\subtable[Average Flux]{
\begin{tabular}{|c|c|c|} 
\toprule%
$\dx$ & Rate($E_h^2$) &  Rate($E_h^{\infty}$) \\ \midrule
0.5          & 1.910 & 1.978 \\ \hline
0.25        & 1.910 & 1.954 \\ \hline
0.125     & 1.923 &  1.937 \\ \hline
0.0625   & 1.936 &  1.941 \\ \hline
0.03125 & 1.946 &  1.948 \\ 
\bottomrule%
\end{tabular}}
\subtable[TVD2 MinMod]{
\begin{tabular}{|c|c|c|} 
\toprule%
$\dx$ & Rate($E_h^2$) &  Rate($E_h^{\infty}$) \\ \midrule
0.5          & 2.042  & 2.032 \\ \hline
0.25        & 2.033  & 2.029 \\ \hline
0.125     & 2.026  &  2.023 \\ \hline
0.0625   & 2.021  &  2.019 \\ \hline
0.03125 & 2.017 &  2.016 \\ 
\bottomrule%
\end{tabular}}
\caption{Rates of convergence.}%
\label{ROC}%
\end{table}

\begin{figure}%
\centering
\subfigure[Evolution of $\eta$ amplitude]{\includegraphics[scale=0.3025]{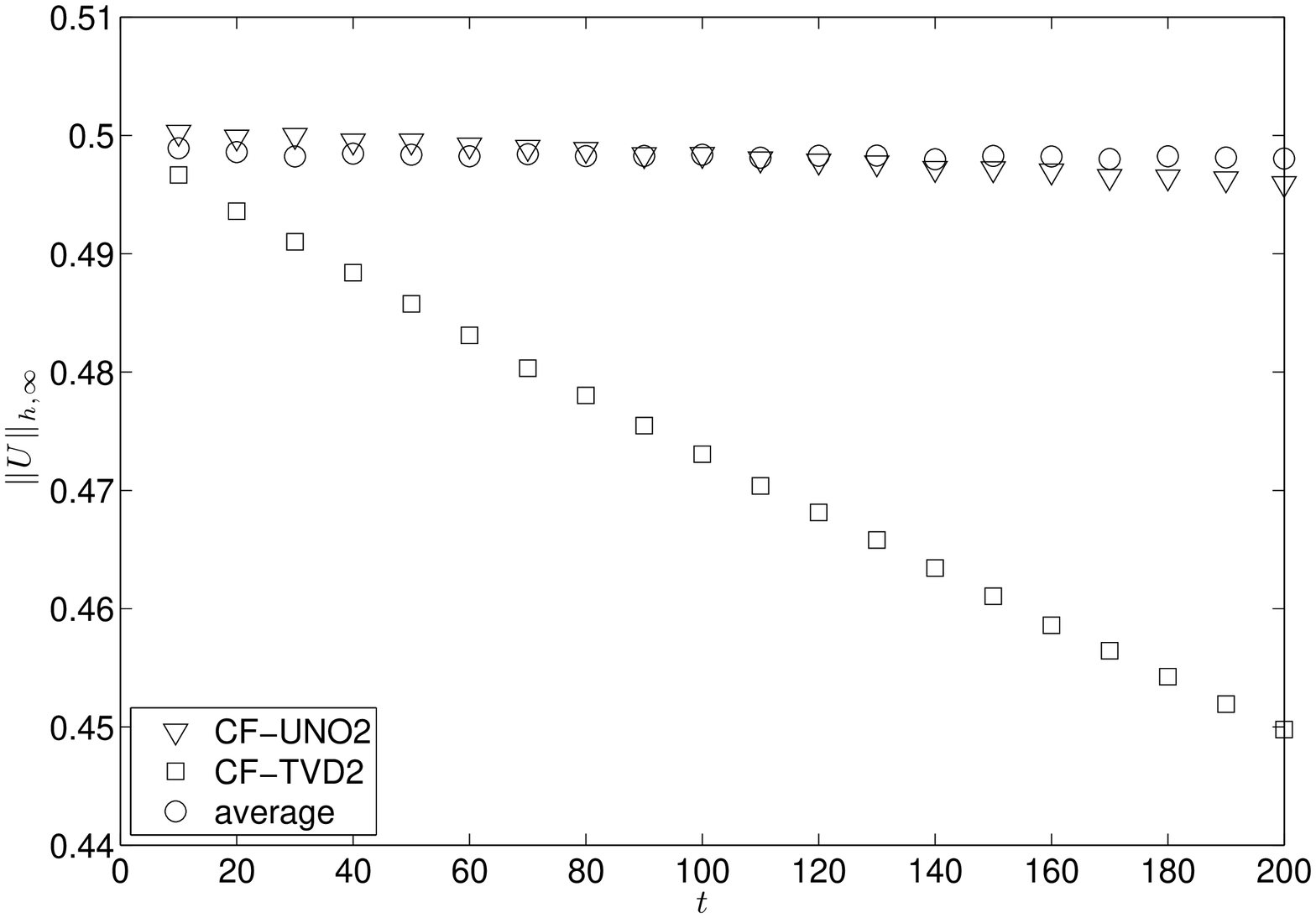}}
\subfigure[Evolution of $I_1^h$]{\includegraphics[scale=0.3025]{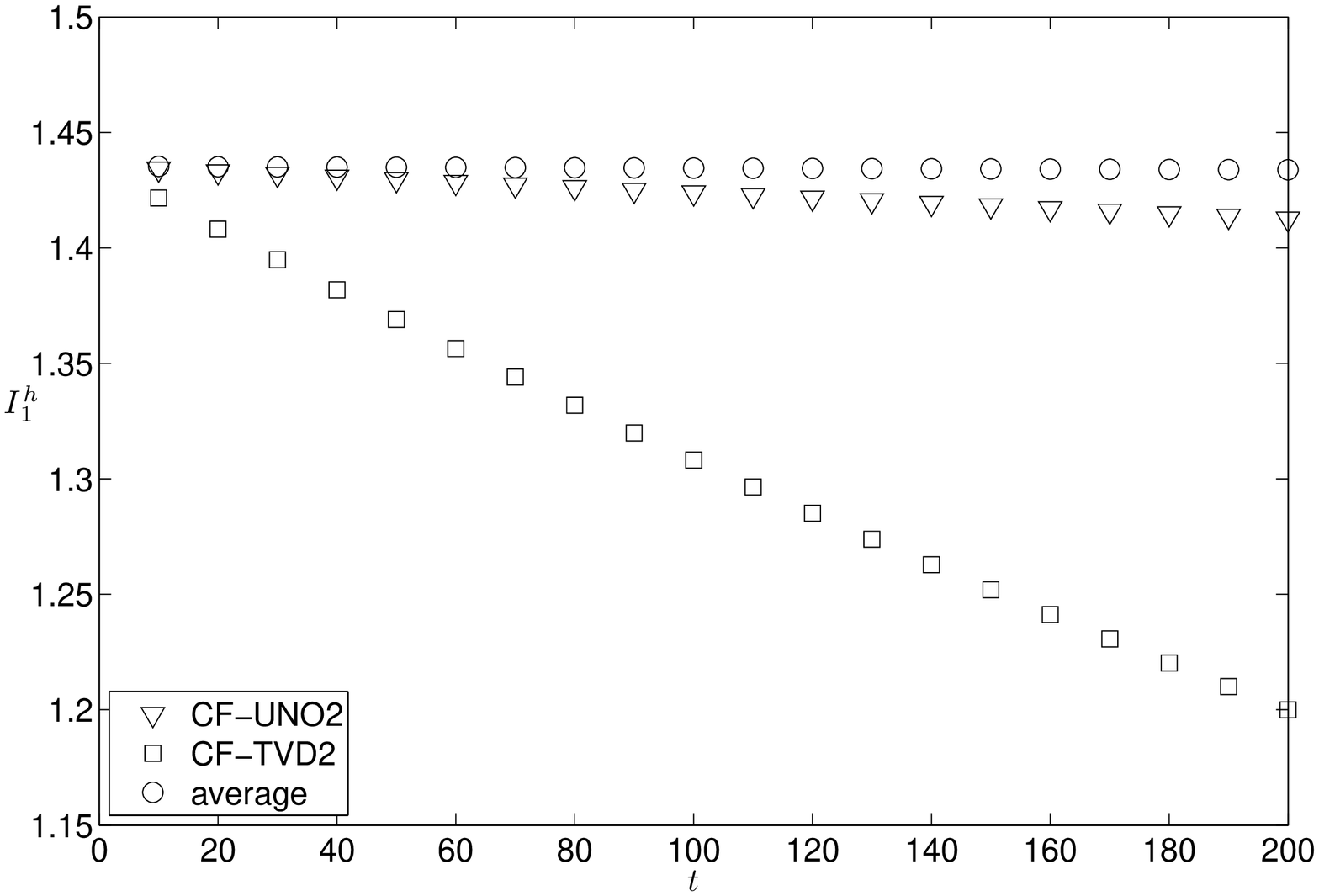}}
\caption{Preservation of the solitary wave amplitude and conservation of the invariant $I_1^h$: $G^{m}$ flux with Minmod limiter}%
\label{F12}%
\end{figure}

We also check the preservation of the invariant $I_1(t)=\int_{\R} (\eta^2(x,t)+(1+\eta(x,t))u^2(x,t)-c\,\eta_x^2(x,t)-a\,u_x^2(x,t))\; dx$ by computing its discrete counterpart:
\begin{equation}\label{EInv2}
I_1^h=\sum_i \dx\left(\eta_i^2+[(1+\eta_i)u_i]^2-c\left[\frac{\eta_{i+1}-\eta_i}{\dx}\right]^2-
a\left[\frac{u_{i+1}-u_i}{\dx}\right]^2\right),
\end{equation}
as well as the discrete mass $I_0^h=\dx\sum_i \eta_i$.  Figure \ref{F12} shows the evolution of the amplitude and the invariant $I_1^h$ of the  solitary wave up to $T=200$. The comparison of various methods is performed. We observe that the UNO2 reconstruction is more accurate  while KT and the CF schemes show comparable performance. We note that the invariant $I_0^h = 1.932183566158$ conserved the digits shown for all numerical schemes. In this experiment we took $\dx=0.1, \ \dt = \dx/2$. 

\subsection{Head-on collisions}

The head-on collision of two counter-propagating solitary waves is characterized by the change of the shape along with a small phase-shift of the waves as a consequence of the nonlinearity and dispersion. These effects have been studied extensively  before by numerical means using high order numerical methods such as finite differences, \cite{BC}, spectral and finite element methods \cite{ADM2} and experimentally in \cite{CGHHS}. In Figure \ref{F15a}  we present the numerical solutions of the BBM-BBM system and the Bona-Smith system with $\theta^2=9/11$ (in dimensional and unscaled variables) along with the experimental data from \cite{CGHHS}. The spatial variable is expressed in centimeters while the time  in seconds. The solutions were obtained using the CF-scheme with UNO2 and WENO3  reconstruction using $\dx = 0.05$ cm and $\dt=0.01$ s. For this experiment we constructed solitary waves for  Boussinesq systems by solving the respective o.d.e's system in the spirit of \cite{Bona2007} such that they fit to experimentally generated solitary waves before the collision. The speeds of the right and left-traveling solitary waves are $c_{r,s}=0.854 \mbox{ m/s}$ and $c_{l,s}=0.752 \mbox{ m/s}$ respectively.


\begin{figure}%
\centering
\subfigure[$t=18.29993 s$]{\includegraphics[scale=.3025]{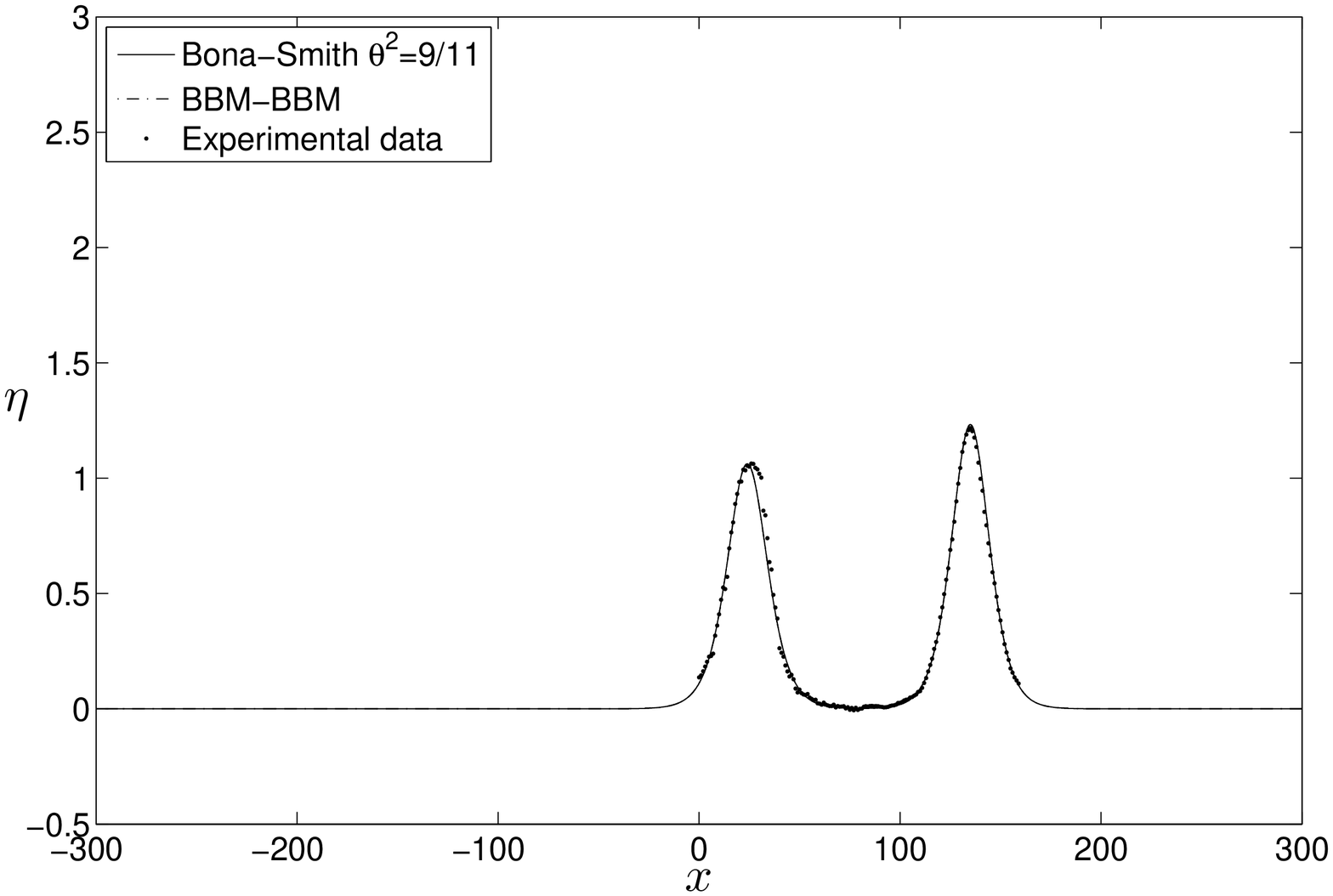}}
\subfigure[$t=18.80067 s$]{\includegraphics[scale=.3025]{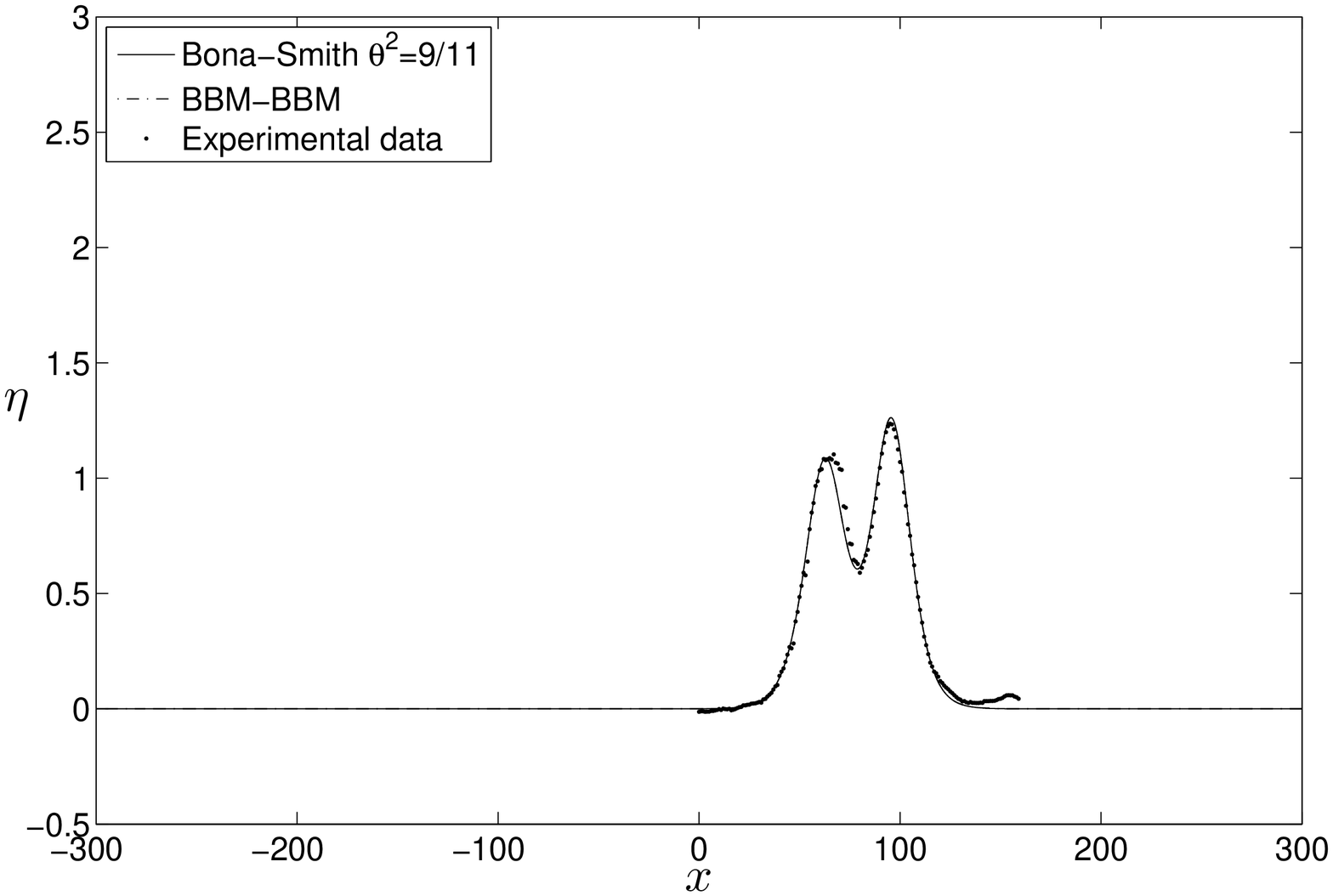}} 
\subfigure[$t=19.00956 s$]{ \includegraphics[scale=.3025]{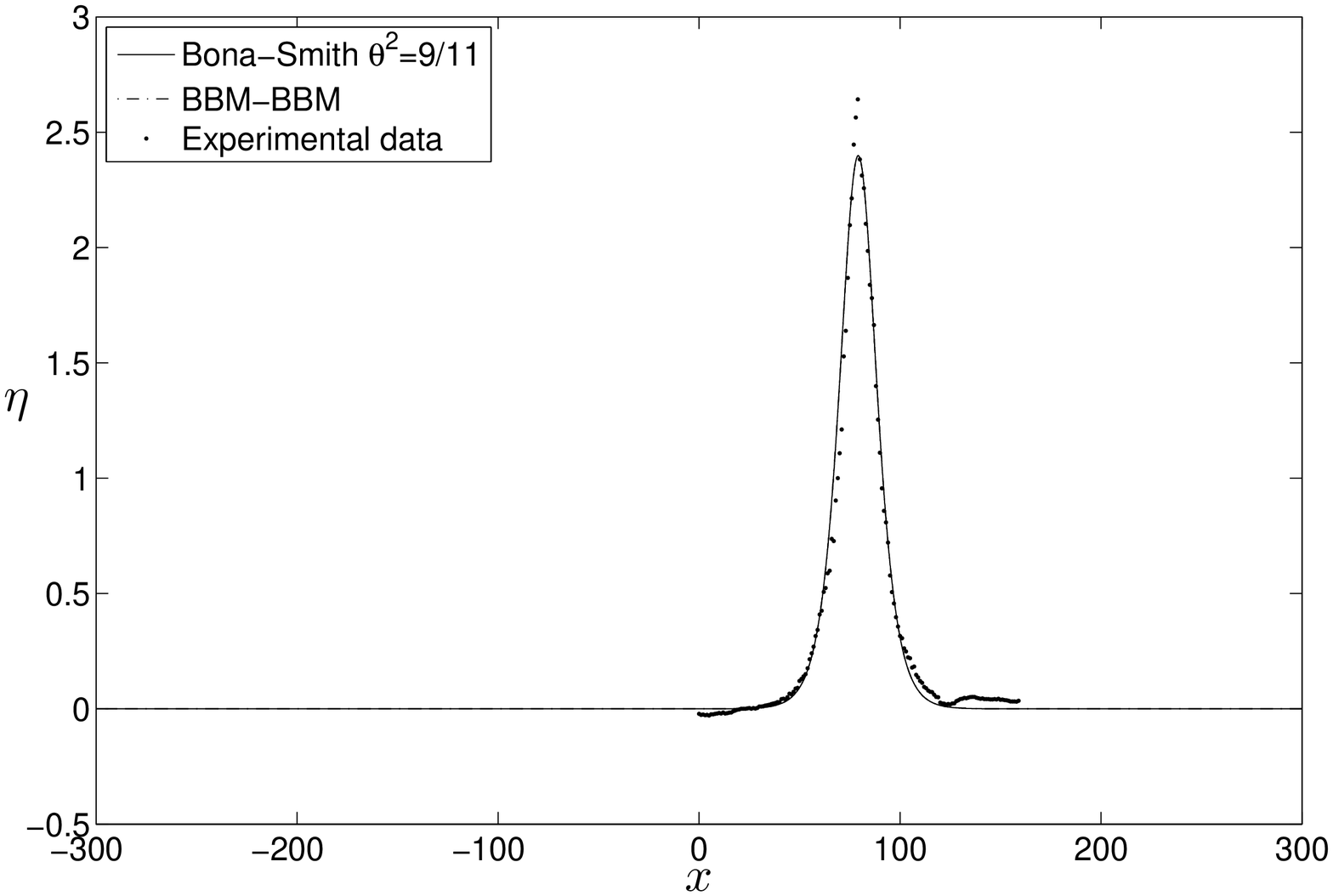}} 
\subfigure[$t=19.15087 s$]{\includegraphics[scale=.3025]{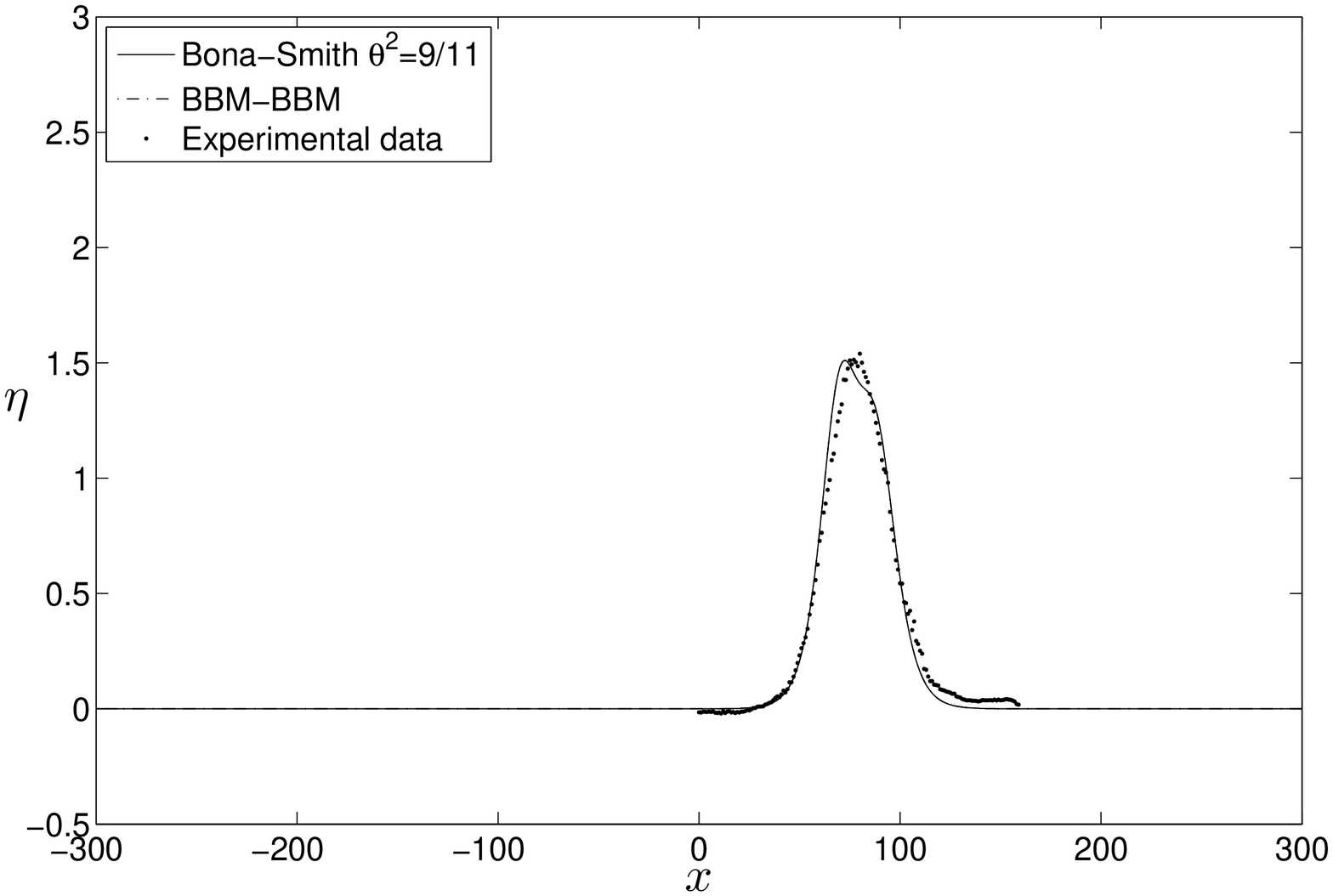}} 
\subfigure[$t=19.19388 s$]{\includegraphics[scale=.3025]{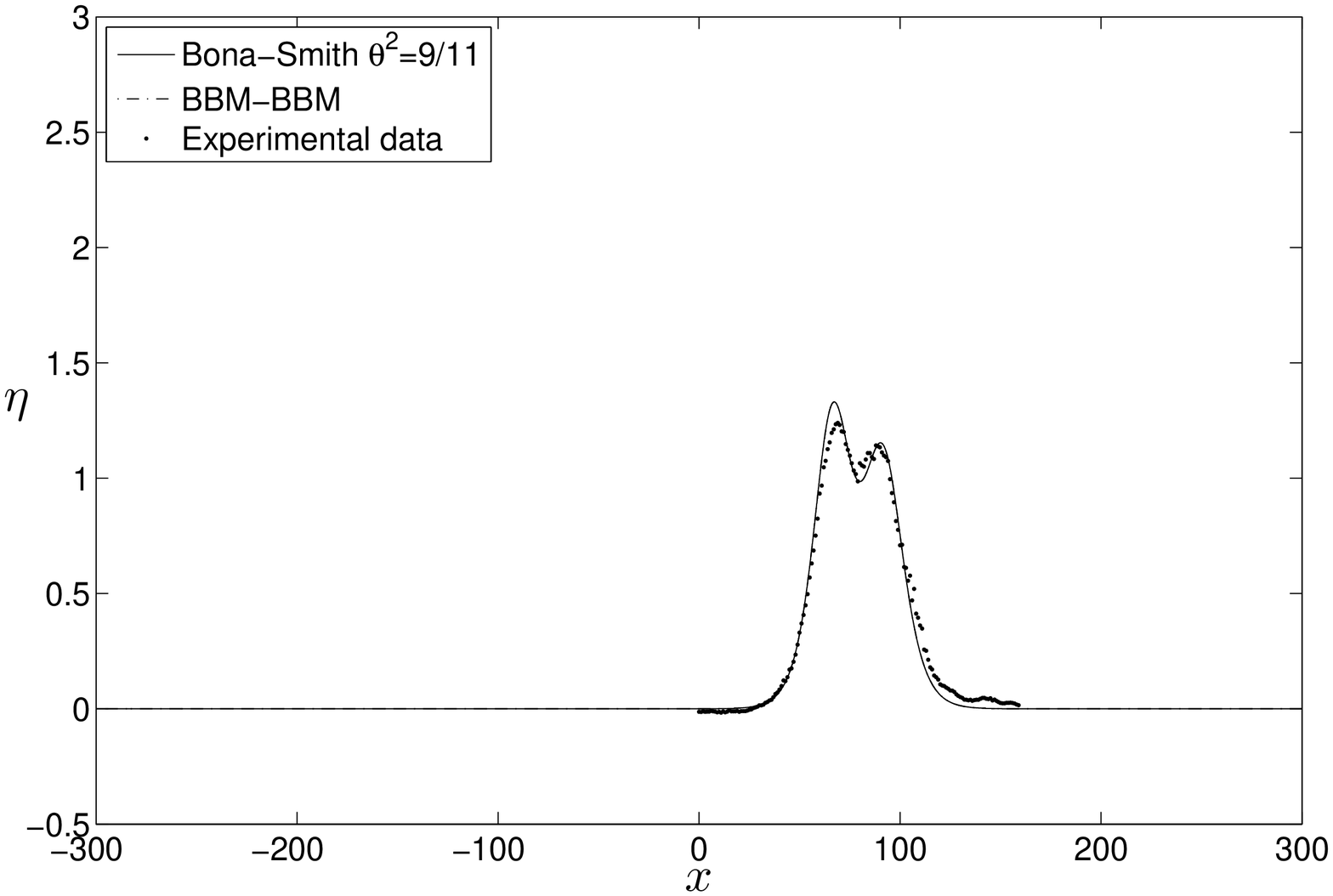}} 
\subfigure[$t=19.32904 s$]{\includegraphics[scale=.3025]{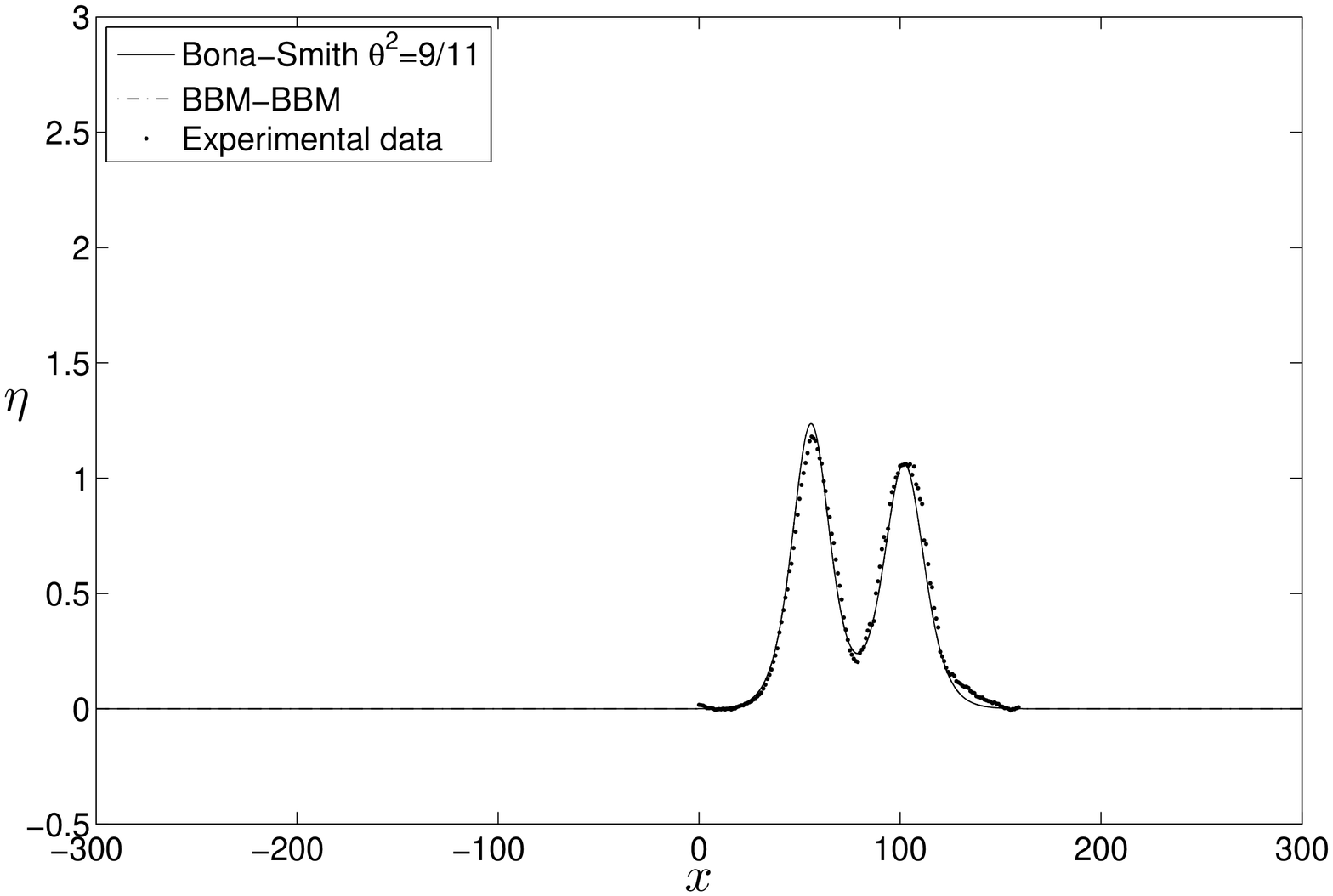}} 
\caption{Head-on collision of two solitary waves: ---: BBM-BBM, $--$: Bona-Smith ($\theta^2=9/11$), \textbullet: experimental data of \cite{CGHHS}}%
\label{F15a}%
\end{figure}


\begin{figure}%
\ContinuedFloat
\centering
\subfigure[$t=19.84514 s$]{\includegraphics[scale=.3025]{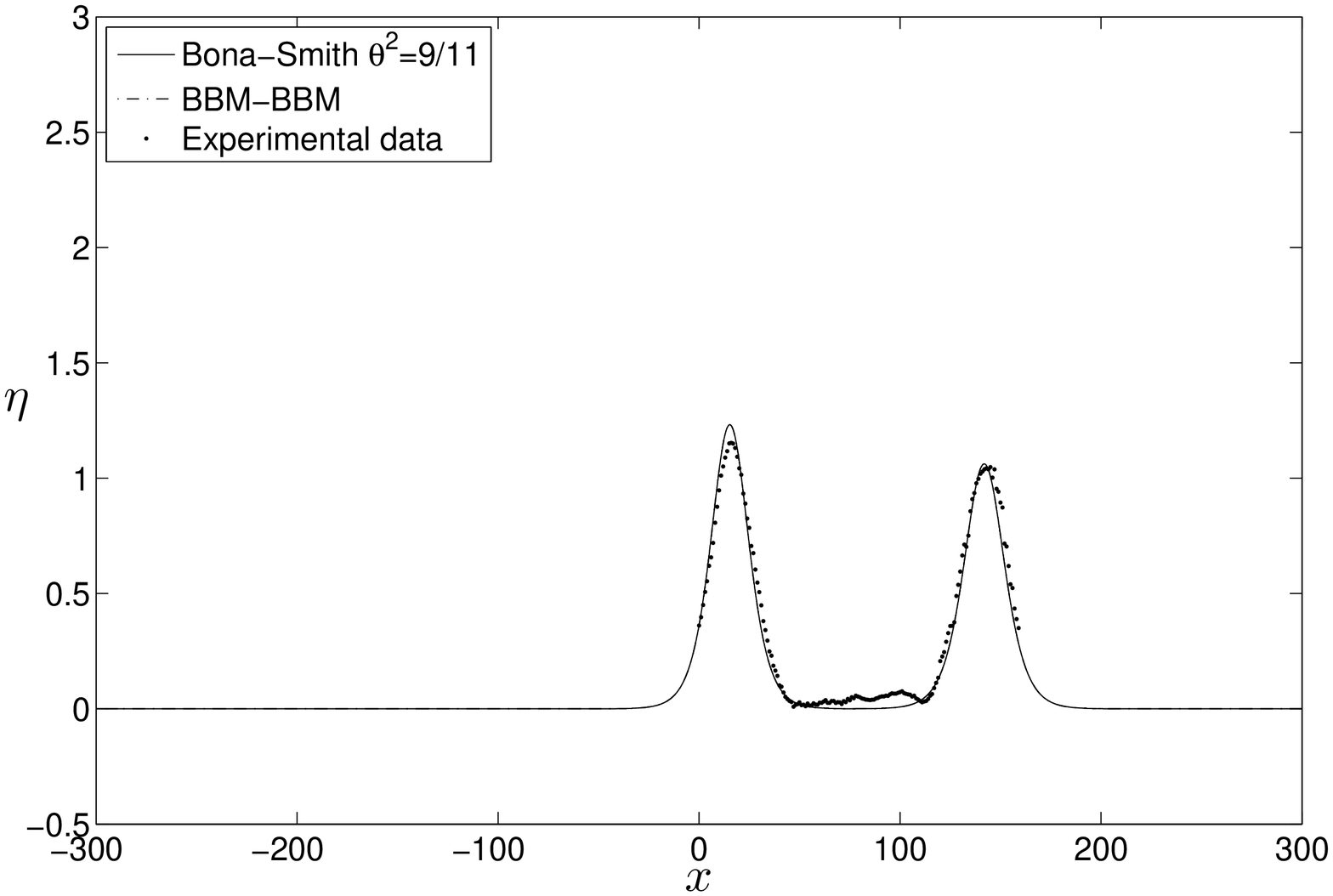}}
\subfigure[$t=20.49949 s$]{\includegraphics[scale=.3025]{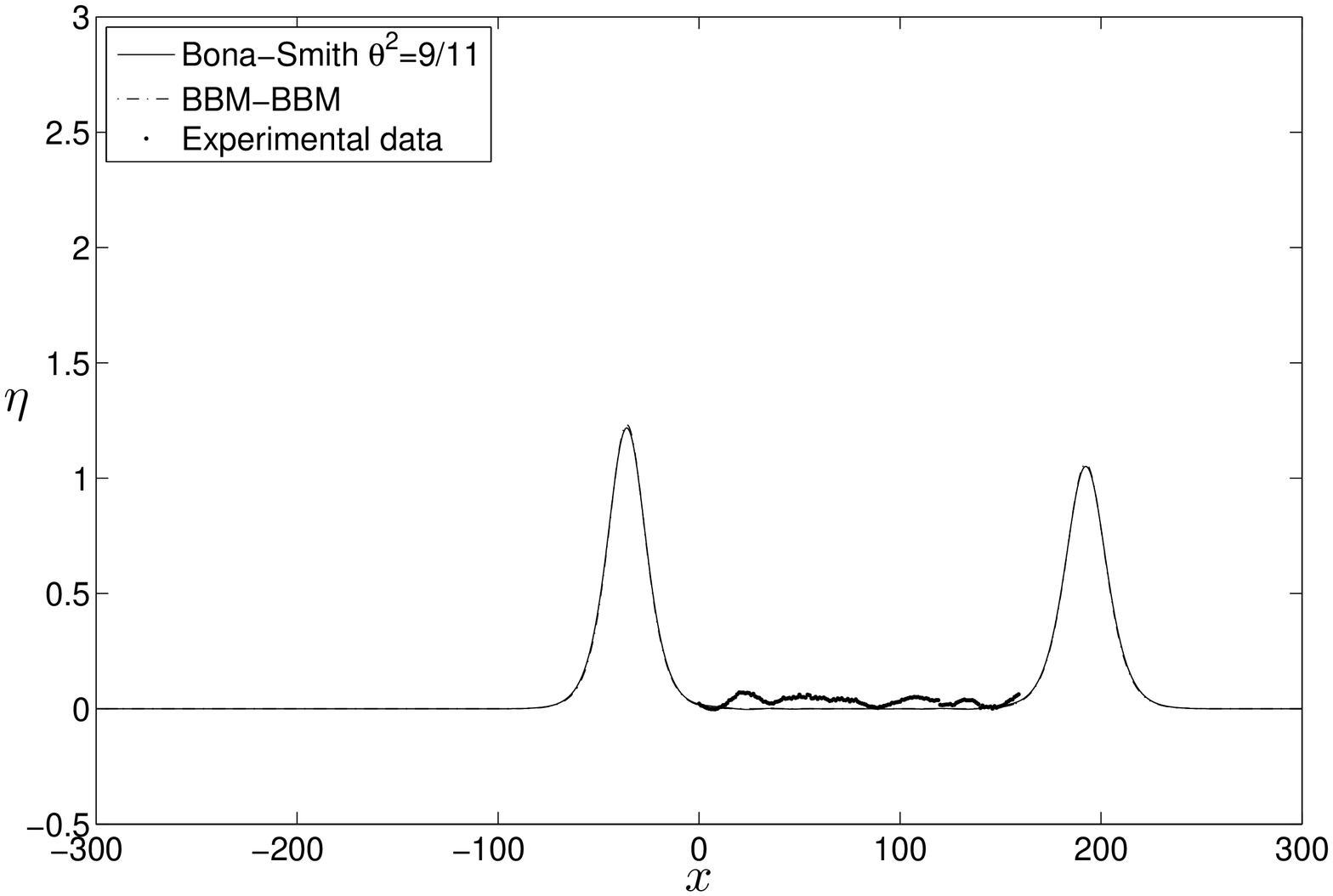}}
\caption{(Cont'd) Head-on collision of two solitary waves. ---: BBM-BBM, $--$: Bona-Smith ($\theta^2=9/11$), \textbullet: experimental data of \cite{CGHHS}}%
\label{F15b}%
\end{figure}

We observe that Boussinesq models converge to the same numerical solution with all numerical schemes we tested. A very good agreement with the experimental data is observed. The discrete mass for the Bona-Smith system is $I_0^h = 0.0059904310418$  and for the BBM-BMM system is $I_0^h = 0.0059199389479$ for all fluxes and reconstructions used. The variances in $I_1^h$ are mainly due to different types of reconstruction and not to the choice of numerical fluxes. In Table \ref{TINV} these values are reported.

\begin{table}%
\centering
\subtable[Bona-Smith]{
\begin{tabular}{|c|l|} 
\toprule%
           &   $I_1^h$ \\ \midrule
m-flux     &   0.000944236 \\ \hline
UNO2       &   0.00094423 \\ \hline
TVD2       &   0.00094 \\ \hline
WENO3      &   0.00094423\\ 
\bottomrule%
\end{tabular}}
\subtable[BBM-BBM]{
\begin{tabular}{|c|l|} 
\toprule%
           &   $I_1^h$ \\ \midrule
m-flux     &   0.00092793 \\ \hline
UNO2       &   0.00092793 \\ \hline
TVD2       &   0.00092 \\ \hline
WENO3      &   0.00092793\\ 
\bottomrule
\end{tabular}}
\caption{Preservation of the invariant $I_1^h$.}
\label{TINV}
\end{table}

\section{Conclusions}\label{sec:concl}

Initially, the finite volume method was proposed by S. Godunov \cite{Godunov1999} to compute approximate solutions to hyperbolic conservation laws. In the present study we made a further attempt to generalize this method to the framework of dispersive PDEs. This type of equations arises naturally in many physical problems. In the water wave theory dispersive equations have been well known since the pioneering work of J. Boussinesq \cite{bouss} and Korteweg-de Vries \cite{KdV}. Currently, the so-called Boussinesq-type models become more and more popular as an operational model for coastal hydrodynamics and other fields of engineering. 

We extend the finite volume framework to dispersive models. We tested several choices of numerical fluxes (average, Kurganov-Tadmor, characteristic), various reconstruction methods ranging from classical (TVD2, UNO2) to modern approaches (WENO3, WENO5). Various choices of limiters have been also tested out. Advantages of specific methods are discussed and some recommendations are outlined.

\section*{Acknowledgment}

D.~Dutykh acknowledges the support from French Agence Nationale de la Recherche, project MathOcean (Grant ANR-08-BLAN-0301-01) and Ulysses Program of the French Ministry of Foreign Affairs under the project 23725ZA. The work of Th.~Katsaounis was partially supported by European Union FP7 program Capacities (Regpot 2009-1), through ACMAC (http://acmac.tem.uoc.gr). The work of D.~Mitsotakis was supported by Marie Curie Fellowship No. PIEF-GA-2008-219399 of the European Commission. We would like to thank also Professors Diane Henderson and Costas Synolakis for providing us their experimental data and Profs Jerry Bona and Vassilios Dougalis for very helpful discussions.

\bibliographystyle{plain}
\bibliography{biblio}

\end{document}